\pdfoutput=1
\documentclass[aoas,preprint]{imsart}

\RequirePackage[OT1]{fontenc}
\RequirePackage{amsthm,amsmath}
\RequirePackage{natbib}
\RequirePackage[colorlinks,citecolor=blue,urlcolor=blue]{hyperref}
\usepackage{graphicx}

\arxiv{arXiv:2157893}

\startlocaldefs
\numberwithin{equation}{section}
\theoremstyle{plain}

\endlocaldefs

\begin{document}

\begin{frontmatter}
\title{Combining Satellite Imagery and Numerical Model Simulation to Estimate Ambient Air Pollution: An Ensemble Averaging Approach}
\runtitle{Combining Satellite Imagery and Numerical Model Simulation }

\begin{aug}
\author{\fnms{Nancy} \snm{Murray}\thanksref{m1}\ead[label=e1]{nancy.murray@emory.edu}},
\author{\fnms{Howard} \snm{Chang}\thanksref{m1}\ead[label=e2]{howard.chang@emory.edu}},
\author{\fnms{Heather} \snm{Holmes}\thanksref{m2}\ead[label=e3]{hholmes@unr.edu}}
\and
\author{\fnms{Yang} \snm{Liu}\thanksref{m1}
\ead[label=e4]{yang.liu@emory.edu}}

\runauthor{N. Murray et al.}

\affiliation{Emory University\thanksmark{m1} and University of Nevada, Reno\thanksmark{m2}}

\address{N. Murray\\
H. C. Chang\\
Department of Biostatistics and Bioinformatics
\\
Emory University
\\
Atlanta, Georgia 30322
\\
USA\\
\printead{e1}\\
\phantom{E-mail:\ }\printead*{e2}}

\address{H. Holmes\\
Department of Physics\\
University of Nevada, Reno\\
Reno, Nevada 89557\\
\printead{e4}}

\address{Y. Liu\\
Department of Environmental Health
\\
Emory University
\\
Atlanta, Georgia 30322
\\
USA\\
\printead{e4}}

\end{aug}

\begin{abstract}
Ambient fine particulate matter less than 2.5 $\mu$m in aerodynamic diameter (PM$_{2.5}$) has been linked to various adverse health outcomes and has, therefore, gained interest in public health. However, the sparsity of air quality monitors greatly restricts the spatio-temporal coverage of PM$_{2.5}$ measurements, limiting the accuracy of PM$_{2.5}$-related health studies. We develop a method to combine estimates for PM$_{2.5}$ using satellite-retrieved aerosol optical depth (AOD) and simulations from the Community Multiscale Air Quality (CMAQ) modeling system. While most previous methods utilize AOD or CMAQ separately, we aim to leverage advantages offered by both methods in terms of resolution and coverage by using Bayesian model averaging. In an application of estimating daily PM$_{2.5}$ in the Southeastern US, the ensemble approach outperforms statistical downscalers that use either AOD or CMAQ in cross-validation analyses. In addition to PM$_{2.5}$, our approach is also highly applicable for estimating other environmental risks that utilize information from both satellite imagery and numerical model simulation.
\end{abstract}


\end{frontmatter}
%%%%%%%%%%%%%%%%%%%%%%%%%%%%%%%%%%%%%%%%%%%%%%%%%%%%%%%%%%%%%%%%%%%%%%%%

%%%%%%%%%%%%%%%%%%%%%%%%%%%%%%%%%%%%%%%%%%%%%%%%%%%%%%%%%%%%%%%%%%%%%%%%
\section{Introduction}
%%%%%%%%%%%%%%%%%%%%%%%%%%%%%%%%%%%%%%%%%%%%%%%%%%%%%%%%%%%%%%%%%%%%%%%%
%no comma with et al.
Air pollution negatively impacts human health, as supported by various studies around the world [\cite{lancet_ap}; \cite{Nit_diox_interest};  \cite{ozone_China};  \cite{NO2};  \cite{CO_interest}; \cite{CV_PM_all}]. %\cite{CV_PM_all, lancet_ap,ozone_China,NO2, Nit_diox_interest, CO_interest}. 
While air pollution represents a complex mixture of chemicals, particulate matter less than $2.5$ $\mu$m in aerodynamic diameter (PM$_{2.5}$), in particular, has received increasing interest in the public health realm [\cite{pm_China}; \cite{pm_in_US}; \cite{PM_India}].
%\cite{pm_China, pm_in_US, PM_India}. 
PM$_{2.5}$ is a mixture of solids and liquids that can penetrate deep into the lower respiratory system to affect the lungs and circulatory system [\cite{Circ}; \cite{Circ_Spain}; \cite{Lung}].
%\cite{Lung, Circ, Circ_Spain}.  
Also, sources of PM$_{2.5}$ include power generation, industrial operations, automobiles; other sources include wildfires, wind blown dust, and ocean spray. Hence, regulatory policies on certain anthropogenic emissions and the changing climate can have noteable impacts on PM$_{2.5}$ concentrations and, subsequently, on human health. As a result, the United States Environmental Protection Agency (USEPA) regulates PM$_{2.5}$ as one of its criteria pollutants to protect public health [\cite{EPA_reg}]. %\cite{EPA_reg}.

Population-based studies of air pollution and health have contributed significantly to setting regulatory standards worldwide. However, these studies draw criticism due to the routine estimation of exposures from regulatory monitoring networks. Monitors in these networks are preferentially located in specific geographic areas, often in areas with high pollution levels and large populations. Due to high cost of maintenance, PM$_{2.5}$ monitor measurements  are spatially sparse, such that using these measurements over a large spatial domain may be inappropriate, and are sometimes temporally available only in 1-in-3 or 1-in-6 days time periods. More recently, an important research area in environmental engineering and epidemiology involves developing data fusion products that supplement monitoring measurements with numerical model simulations and remotely-sensed observations from satellites. These data fusion models typically involve hindcasting as a means of supplementing health analyses. The overarching goal of data fusion is to increase the spatial-temporal coverage of air quality data to support health analyses and health impacts assessments. 

Numerical models used in air pollution research are known as chemical transport models (CTM). CTMs are 3-dimensional deterministic models that simulate gridded air pollution concentrations based on state-of-the-art knowledge on drivers of air quality [\cite{Chipperfield}].  Advantages of CTM include its complete spatial-temporal coverage and the ability to incorporate chemical and physical processes associated with air pollution. However, CTM is computationally expensive and often is only available at crude spatial resolutions.
For PM$_{2.5}$, remotely-sensed aerosol optical depth (AOD) has been examined extensively in its ability to predict PM$_{2.5}$ in combination with other meteorological and land use variables [\cite{Liu_US}; \cite{Liu_2009}]. %\cite{Liu 2009, SE US remote sensing}.
AOD measures the degree to which aerosols prevent light from penetrating the atmosphere. Some main advantages of satellite-based AOD are its fine spatial resolution, global coverage, and public accessibility. However, remotely-sensed data can suffer from missing data due to retrieval error and cloud cover. 

CTM simulations and AOD values cannot be used directly in health analyses because complex spatial-temporal bias exists when compared to ground-level monitoring data. For example, the Community Multiscale Air Quality (CMAQ) model, a type of CTM, may suffer from underprediction or overprediction due to error in inputs and discretization over space and time [\cite{Mebust_2003}; \cite{Lim}]. %\cite{Mebust 2003, Lim}. 
AOD measures aerosol over the entire atmospheric column and its relationship with ground-level PM$_{2.5}$ can depend on various factors. Therefore, data fusion models that calibrate CTM and AOD data against observed measurements are needed [\cite{Berrocal_downscaler}; \cite{Chang_JESEE}]. %\cite{Berrocal_downscaler, Chang_JESEE}.

Most existing data fusion models have been developed to utilize only one data source - CTM or satellite AOD.  Concurrent utilization of both data sources in the fusion process may provide more accurate PM$_{2.5}$ estimates. Specifically, CTM simulation can address the missing data problem in satellite AOD, while satellite AOD can provide additional fine-scale spatial information to CTM simulation. CTM simulations, however, are not nudged to observations while AOD retains some form of observed data. Current approaches center around using CTM simulations to impute missing AOD values, followed by using the gap-filled AOD field as a predictor of PM$_{2.5}$ in regression models [\cite{Xiao_2017}] or machine learning algorithms [\cite{Di_2016}]. Similarly, in the Global Burden of Disease project of \cite{Multi_proxy}, annual PM$_{2.5}$ averages are obtained by using satellite AOD values that are informed by CTM simulations to account for the vertical aerosol profile. %\cite{Multi-proxy}.  

In this paper, we describe a Bayesian ensemble approach to perform data fusion with multiple sources of information. Specifically, predictions from data fusion models using either CTM simulation or satellite AOD are combined with spatially-varying weights. Our model-based ensemble approach offers several advantages compared to previous methods, namely the ability to incorporate various sources of uncertainty in predictions and to characterize the relative prediction performance of CTM versus satellite AOD. In an application, we evaluate the proposed ensemble method for predicting daily PM$_{2.5}$ in the Southeastern United States (Southeastern US). 

The remainder of the article is organized as follows. First, we introduce our motivating air quality, numerical model simulation, and remote sensing data in Section 2. In Section 3, we describe the proposed ensemble method and estimation approach under a Bayesian framework. Section 4 presents results of (1) an evaluation of the ensemble approach compared to methods that utilize the CMAQ simulation or AOD separately, (2) estimation of fine-scaled weights across the Southeastern US, and (3) estimation of fine-scaled PM$_{2.5}$ concentrations with complete spatial-temporal coverage for the Atlanta metropolitan area. Finally, we discuss strengths, limitations, and future work of our method in Section 5. 

%%%%%%%%%%%%%%%%%%%%%%%%%%%%%%%%%%%%%%%%%%%%%%%%%%%%%%%%%%%%%%%%%%%%%%%%
\section{Data}
%%%%%%%%%%%%%%%%%%%%%%%%%%%%%%%%%%%%%%%%%%%%%%%%%%%%%%%%%%%%%%%%%%%%%%%%

We obtained daily ground-level 24-hour average measurements of PM$_{2.5}$ from 63 monitors in the Southeastern US  over the period 2003 to 2005 via the USEPA's Air Quality System (AQS). Computer model simulations were obtained from the USEPA Models-3/Community Multiscale Air Quality (CMAQ) model version 4.5 at a 12 km $\times$ 12 km horizontal spatial resolution [\cite{CMAQ}]. %\cite{CMAQ}. 
We acquired satellite-retrieved AOD measurements by the aerosol remote sensor Moderate Resolution Imaging Spectroradiometer (MODIS), which orbits the Earth on the National Aeronautics and Space Administration's Aqua and Terra satellites.  We utilized a new multiangle implementation of atmospheric correction (MAIAC) algorithm that provides AOD values at a 1km $\times$ 1km spatial resolution [\cite{AOD1}; \cite{AOD2}].
%\cite{AOD1,AOD2}. 
For each AOD grid cell, we also compiled variables including: elevation from the US Geological Survey, forest cover and road lengths from the 2001 National Land Cover data, meteorology (e.g. wind speed) from the North American Land Data Assimilation Systems, and PM$_{2.5}$ primary emission point sources from the 2002 USEPA National Emissions Inventory. As in \cite{1kmAOD_Hu}, forest cover and elevation were averaged from their original resolutions of about 1 km and about 30 m, respectively, to the 1 km $\times$ 1 km MAIAC grid cell level. Additionally, road lengths and point emissions were summed over the 1 km $\times$ 1 km MAIAC grid cell level.

\begin{figure}[h]
     \centering
%trim={<left> <lower> <right> <upper>}
\scalebox{.7}{\includegraphics[trim={3cm 13.5cm 1cm 4.1cm}, clip]{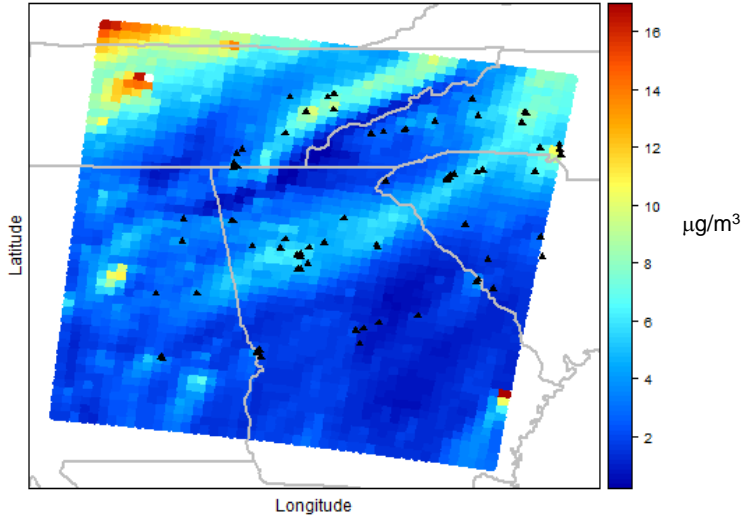}}
\caption{Simulation of PM$_{2.5}$ from the Community Multiscale Air Quality (CMAQ) model on March 17, 2015. Values are plotted at the centroid of each 12 km $\times$ 12 km grid cell.}
\label{fig:CMAQ}
 \end{figure}

\begin{figure}[h]
     \centering
\scalebox{.7}{\includegraphics[trim={2.5cm 13.7cm 5cm 4.2cm}, clip]{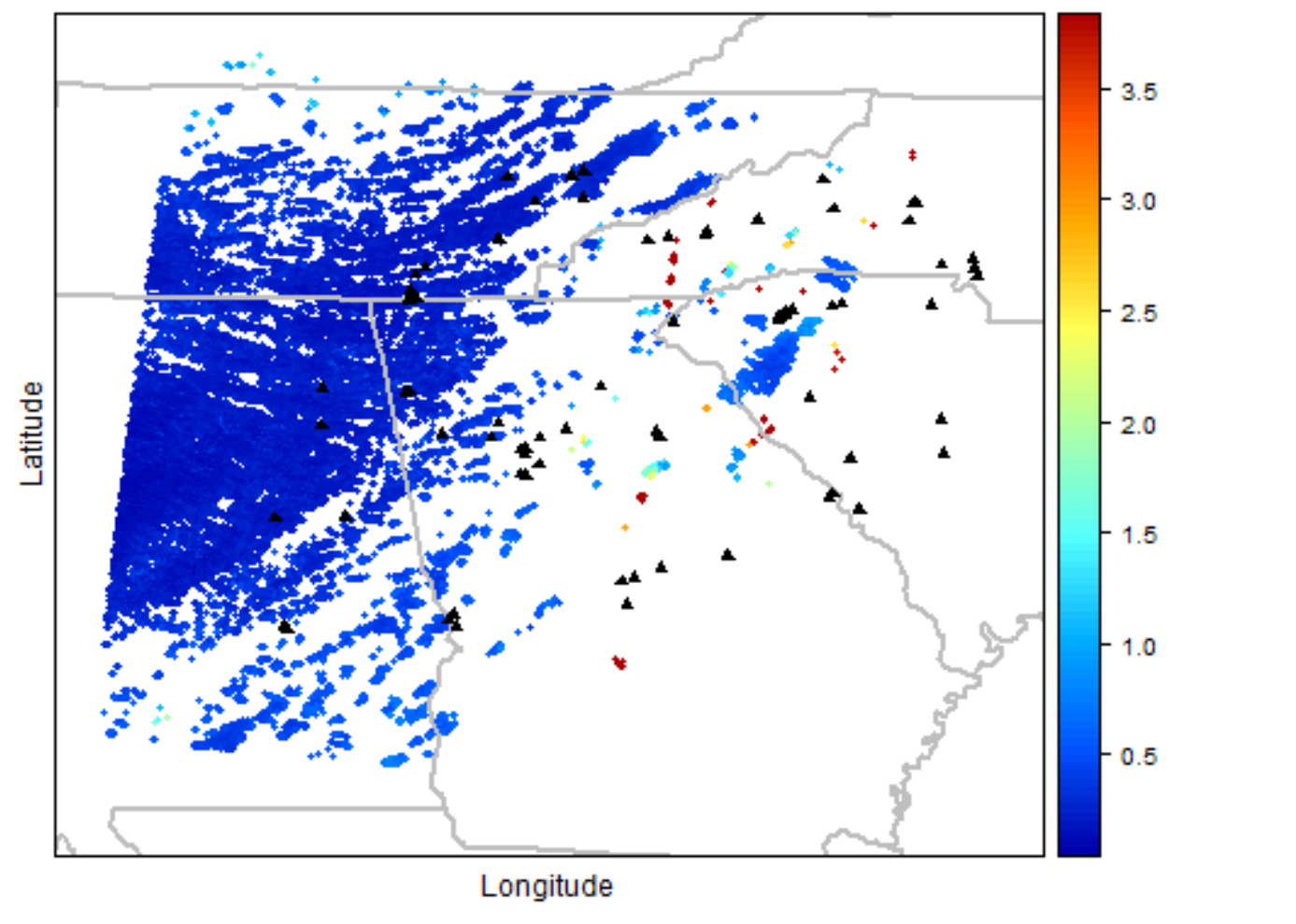}}
\caption{Satellite-derived aerosol optical depth (AOD) 1 km $\times$ 1 km gridded resolution on March 17, 2015.}
\label{fig:AOD}
 \end{figure}

Figure \ref{fig:CMAQ} shows the locations of the 63 AQS monitors in our study region and gridded PM$_{2.5}$ simulations from CMAQ on March 17th, 2015. Similarly, Figure \ref{fig:AOD}, with an overlay of the same AQS monitor locations, shows the 1km-level satellite MAIAC AOD values on the same day with a considerable amount of missing data. 

%CMAQ figure here
 
 %AOD figure here

%A value of 0.01 corresponds to an extremely clean atmosphere, and a value of 0.4 would correspond to a very hazy condition. An average aerosol optical depth for the U.S. is 0.1 to 0.15. (https://www.esrl.noaa.gov/gmd/grad/surfrad/aod/)
%%%%%%%%%%%%%%%%%%%%%%%%%%%%%%%%%%%%%%%%%%%%%%%%%%%%%%%%%%%%%%%%%%%%%%%%
%%%%%%%%%%%%%%%%%%%%%%%%%%%%%%%%%%%%%%%%%%%%%%%%%%%%%%%%%%%%%%%%%%%%%%%%
\section{Methods}
%%%%%%%%%%%%%%%%%%%%%%%%%%%%%%%%%%%%%%%%%%%%%%%%%%%%%%%%%%%%%%%%%%%%%%%%
%We introduce previously established methods, the statistical downscaler and Bayesian Model Averaging (BMA), before describing the proposed ensemble model and its estimation procedure.
%%%%%%%%%%%%%%%%%%%%%%%%%%%%%%%%%%%%%%%%%%%%%%%%
\subsection{Spatial-Temporal Statistical Downscaling}
\label{sec:downscaler}
%%%%%%%%%%%%%%%%%%%%%%%%%%%%%%%%%%%%%%%%%%%%%%%%
%Conventions when referring to the downscaler: 
%Downscaling Model
%Downscaler 
We first describe the model for combining monitoring data with CMAQ or AOD. 
We take a statistical downscaling approach that treats gridded CMAQ outputs or AOD values as predictors for point-referenced AQS monitoring measurements in a Bayesian spatial hierarchical model. Predictions of PM$_{2.5}$ from the AOD and CMAQ downscalers are subsequently used as inputs to the ensemble model. 

Let $Y_{st}$ represent the observed PM$_{2.5}$ concentration on day $t$ at location $s$. Following \cite{Berrocal_downscaler} and \cite{Chang_JESEE}, %\cite{Berrocal_downscaler, Chang_JESEE}
the downscaler model has the form: 
\begin{equation}\label{eq:downscaler}
Y_{st} = \alpha_{st} + \beta_{st}X_{st} + \textbf{Z}_{st}\mbox{\boldmath$\gamma$} + \epsilon_{st},
\end{equation}
where $X_{st}$ is the linked AOD or CMAQ value in the grid cell containing the monitor at location $s$, and  $\textbf{Z}_{st}$ is a vector of additional predictors with coefficient {\boldmath $\gamma$}. For the AOD model, $\textbf{Z}_{st}$ includes the following land use and meteorology variables: elevation, forest cover, road length, primary emission source, wind speed, and temperature.  %Nancy double check
Because CMAQ uses information on emissions and meteorology to perform simulations, $\textbf{Z}_{st}$ is not included in the CMAQ downscaler. Preliminary analysis also showed that including additional covariates does not improve prediction performance for the CMAQ model. Finally, the residual error term, $\epsilon_{st} $, is independent normally distributed with mean zero and variance $\sigma_y^{2}$. %\distas{iid} N(0,\sigma^{2}) $.
        
Parameters $\alpha_{st}$ and $\beta_{st}$ in Equation (\ref{eq:downscaler}) are spatial-temporal random effects, sometimes referred to as calibration parameters because they correct for the additive and multiplicative bias associated with CMAQ or AOD. We assume the spatial and temporal random effects are additive: $\alpha_{st} = \alpha_{0,t} + \alpha_{1,s} $ and $\beta_{st} = \beta_{0,t} + \beta_{1,s}$. Temporal dependence in $\alpha_{0,t}$ and $\beta_{0,t}$ is modeled using a first-order conditional autoregressive model (CAR). The CAR model is defined via temporal adjacencies. Let $t\sim t'$ indicate that days $t$ and $t'$ are 1 day apart. The full conditional distribution of $\alpha_{0,t}$ is Gaussian with $E[\alpha_{0,t}] = \eta_{\alpha_0} \sum_{t'\sim t}\alpha_{0, t'}/n_t$ and  $\text{Var}[\alpha_{0,t}] = \sigma_{\alpha_0}^2/n_t$, where $n_t$ is the number of temporal neighbors and  $\eta_{\alpha_0} \in [0,1]$ controls the degree of temporal dependence. Temporal random effects $\beta_{0,t}$ for CMAQ or AOD are defined similarly. Spatial dependence in $\alpha_{1,s}$ and $\beta_{1,s}$ is modeled jointly using a linear coregionalization model. Specifically, we assume  $(\alpha_{1,s}, \beta_{1,s})^T= \textbf{A}\textbf{v}_s$, where $\textbf{A}$ is a $2 \times 2$ lower triangular matrix, and $\textbf{v}_s$ is a $2 \times 1$ vector $(v_{1s}, v_{2s})^T$, where $v_{1s}$ and $v_{2s}$ represent two latent independent Gaussian processes with marginal variances of 1 and exponential covariance functions with range parameters $\theta_j$, i.e. $\text{Cov} (v_{1s}, v_{2s'}) = e^{-||s-s'||/\theta_j}$ for $j = 1, 2$.

%%%%%%%%%%%%%%%%%%%%%%%%
\subsection{Ensemble Modeling}
%%%%%%%%%%%%%%%%%%%%%%%%
Our proposed ensemble method is based on the Bayesian Model Averaging (BMA) framework. BMA has been applied to weather forecasting in order to combine forecasts from different numerical weather models. Here, we extend the approach for estimating spatial-temporal air pollution concentrations when predictions from multiple statistical models are available. BMA provides probabilistic forecasts and, following \cite{Raftery}, is represented as 
%Equation 1 and 2 from Research_Update with Yang (Maybe add the BMA predictive mean too)
\begin{equation}\label{eq:BMA}
        p(y) =  \sum_{k=1}^{K}p(y \mid M_{k})\,p(M_{k} \mid \tilde{y}),
\end{equation}
where $y$ denotes the value to be forecasted ; $M_{1}, ... , M_{K}$ are the $K$ forecast models; $p(y \mid M_{k})$ is the forecast probability density function based on $M_{k}$ alone, and $p(M_{k} \mid \tilde{y})$ is the posterior probability of model $M_{k}$ given training data $\tilde{y}$, with the constraint that $\sum_{k=1}^{K}p(M_{k} \mid \tilde{y}) = 1$.

Let $M_k$ denote the statistical downscaler model for CMAQ ($k =1$) or AOD ($k =2$). We extend the above BMA forecast framework by considering the following model
\begin{equation}\label{eq:Ensemble}
        p(y_{st} \mid M_{1}, M_{2}) =  w_sf_{1}(y_{st}\mid M_{1}) + (1-w_s)f_{2}(y_{st}\mid M_{2}),
\end{equation}
where ${y}_{st}$ is the PM$_{2.5}$ value; $f_{k}(y_{st}\mid M_{k})$ is the posterior \emph{predictive} distribution of ${y}_{st}$ from model $M_k$, and $w_{s}$ is the weight for the CMAQ downscaler at location s.  

Equation (\ref{eq:Ensemble}) can be viewed as a mixture model, where $w_s$ is the posterior probability (ensemble weight) that the downscaler estimate using CMAQ is the best estimate of PM$_{2.5}$ at monitor $s$. Here we assume $f_{k}(y_{st}\mid M_{k})\equiv \phi\left(y_{st} \mid \mu_{st}^{(k)}, \sigma_{st}^{{2,(k)}}\right)$, i.e. a Normal  posterior predictive distribution of ${y}_{st}$ with mean $\mu_{st}^{(k)}$ and variance $ \sigma_{st}^{{2,(k)}}$ using either the CMAQ downscaler ($k=1$) or the AOD downscaler ($k=2$). Hence, the point prediction of $y_{st}$ can be defined by its posterior mean $\hat {y}_{st} = w_{s}\mu_{st}^{(1)}  + (1-w_{s})\mu_{st}^{(2)}$, which is a weighted average of predictions from the CMAQ and the AOD downscalers. The posterior interval can be defined as the 2.5\% and the 97.5\% interval of the mixture distribution. To allow for spatial interpolation of the ensemble weight to locations without monitors, we further assume that $ q_s = \text{logit} (w_{s})$ is a Gaussian process with an exponential covariance function, i.e. $\text{Cov} (q_{s}, q_{s'}) = \tau^2 e^{-||s-s'||/\rho}$.

%%%%%%%%%%%%%%%%%%%%%%%%
\subsection{Estimation and Prediction}
%%%%%%%%%%%%%%%%%%%%%%%%

Estimation and prediction are accomplished in three stages. First, we fit the CMAQ and the AOD downscalers to obtain posterior predictive means, $\mu_{st}^{(k)}$, 
and variances for each observed PM$_{2.5}$ value. To avoid overfitting, each observation was left-out and back-predicted in a \emph{cross-validation} experiment, similar to approaches employed in stack regression [\cite{stacked_reg}]. Second, we fit the ensemble model with spatially-varying weights using the posterior predictive distributions from Stage 1 as inputs. Finally, the CMAQ and the AOD downscalers are fitted again with all PM$_{2.5}$ observations. Predictions are made at all locations and then combined using ensemble weights from Stage 1.

Inference is carried out in a Bayesian framework by specifying priors for all model parameters. For the downscalers (Section \ref{sec:downscaler}), each component of the fixed effect $\mbox{\boldmath$\gamma$}$ is assigned a flat prior ($\propto 1$), and each element of $\textbf{A}$ is assigned $N(0, 1\times10^3)$. The downscaler's temporal CAR parameters $\eta_{\alpha_0}$ and $\eta_{\beta_0}$ are discretized into 1,000 intervals in $[0,1]$. Variance components (marginal variances for the Gaussian process, $\tau^2$; downscaler's residual error variance, $\sigma^2_y$; conditional variance of the temporal CAR model, $\sigma_{\alpha_0}^2$ and $\sigma_{\beta_0}^2$) are assigned Inverse-Gamma $(a, b)$, with $a$ and $b$ chosen to be small and non-informative. Range parameters for Gaussian processes were assigned Gamma (0.5, 0.005).
Markov chain Monte Carlo (MCMC) methods are used to obtain samples from posterior distributions; we use Gibbs sampler when the full conditional distributions are in closed-form and the random-walk Metropolis-Hasting algorithm otherwise. MCMC computations for the downscaler are standard for Bayesian hierarchical modeling and are provided elsewhere [\cite{Chang_JESEE}]. Here we only present the MCMC algorithm for the ensemble model.  

We first introduce a latent variable $z_{st}$, where $z_{st} = 1$ if the prediction from CMAQ performs superiorly to AOD. After initialization, we update $z_{st}$, $w_s$, $\tau^2$, and $\rho$ as follows. 

%\textit{Begin Joint Approach.}
\begin{enumerate}

    \item Update $z_{st}$ for $s = 1, \ldots, S$ and $t = 1, \ldots, T_s$. The full conditional distribution of $z_{st} \sim \text{Bernoulli}$ with probability  
            \[     \frac{ w_s*\phi\left(y_{st} \mid \mu_{st}^{(1)}, \sigma^{2,(1)}_{st}\right)  } {  w_s*\phi\left(y_{st} \mid \mu_{st}^{(1)}, \sigma^{2,(1)}_{st}\right) +   (1-w_s)*\phi\left(y_{st} \mid \mu_{st}^{(2)}, \sigma^{2,(2)}_{st}\right) } \;.
                \]  %4
   \item Update $w_s$ for $s = 1, \ldots, S$.  %which is logit-normally distributed with mean 0 and standard error $S$, where $S[i,j] = \tau^{2} * exp (-1/\rho * d_{ij})$. $d_{ij}$ is the Euclidean distance between $s_{i}$ and $s_{j}$ for all locations i,j.   
   At the $r^{\text{th}}$ iteration, generate a proposal $q_{s}^{(r)}=\text{logit}(w_s^{(r)})$ from a Normal distribution with mean $q_{s}^{(r-1)}$ and variance $\kappa_w$. Accept $q_{s}^{(r)}$ with probability 
   \[
   \frac{\text{logit}(\sum_{t=1}^{T_s}(z_{st}*q_{s}^{(r)}) - {T_s}*\text{log}(1+\text{exp}(q_{s}^{(r)})) h_{1}(q_{s}^{(r)} \mid \tau^2, \rho, \textbf{q}_{-s})}{\text{logit}(\sum_{t=1}^{T_s}(z_{st}*q_{s}^{(r-1)}) - {T_s}*\text{log}(1+\text{exp}(q_{s}^{(r-1)}))h_{1}((q_{s}^{(r-1)} \mid \tau^2, \rho, \textbf{q}_{-s})} ,
   \]
   where $\text{logit}(\sum_{t=1}^{T_s}(z_{st}*q_{s}^{(r)}) - {T_s}*\text{log}(1+\text{exp}(q_{s}^{(r)}))$ is the log-likelihood of the Bernoulli distribution given $q_{s}$, and $h_{1}$($\cdot$) is the univariate conditional Normal distribution given $\textbf{q}_{-s}$, the vector of logit weights for all locations except location $s$.
   %5
   \item Update $\tau^{2}$. The full conditional distribution of $\tau^2 \sim$ Inverse-Gamma$(a + S/2, b + 1/2\textbf{W}^{T}\Sigma^{-1}\textbf{W})$, where $\textbf{W} = (q_1, \ldots, q_S)^T$, and $\Sigma$ is the spatial covariance matrix of $\textbf{W}$. 
    \item
    Update $\rho$.
    Generate proposal $\rho^{(r)}$ from a log-normal distribution with mean $\rho^{(r-1)}$ and variance $\kappa_{\rho}$. Accept $\rho^{(r)}$ with probability 
    \[
    \frac{l_{1}(\textbf{W} \mid \tau^{2}, \rho^{(r)} )\,l_{2}(\rho^{(r)}) \rho^{(r)} }{l_{1}(\textbf{W} \mid \tau^{2}, \rho^{(r-1)} )\,l_{2}(\rho^{(r-1)}) \rho^{(r-1)}},
    \]
    where $l_{1}$ is the multivariate Normal distribution for the Gaussian process, and $l_{2}$ is the Gamma prior distribution.

\end{enumerate}

Updating $q_s$, and thereby $w_s$, individually when the number of monitoring locations is large can be computationally demanding. Hence, we also consider a \emph{two-stage} approach. First, the ensemble weight $w_s$ is estimated separately at each location by assuming a prior distribution $w_s \sim \text{Beta} (1,1)$. Bayesian kriging is then applied to the posterior medians of $w_s$ across locations, assuming a similar Gaussian process model as above. Compared to the original \emph{joint} estimation approach, the two-stage approach assumes $q_s$ to be known when performing spatial interpolation.      
%%%%%%%%%%%%%%%%%%%%%%%%%%%%%%%%%%%%%%%%%%%%%%%%%%%%%%%%%%%%%%%%%%%%%%%%
\section{Application to Southeastern US Daily PM$_{2.5}$ Concentrations}
%%%%%%%%%%%%%%%%%%%%%%%%%%%%%%%%%%%%%%%%%%%%%%%%%%%%%%%%%%%%%%%%%%%%%%%%
\linespread{1.8} 
We evaluated the prediction performance of the proposed ensemble approach using two out-of-sample cross-validation (CV) experiments. First, in a 10-fold CV, we randomly divided the dataset into 10 subsets. Repeatedly, we left out each subset (10\% of the data) and used the other 90\% of the data to fit the prediction model. Because data are available at each monitor in each CV fold, this 10-fold CV experiment allowed us to evaluate the model's ability to perform \emph{temporal} interpolation when daily PM$_{2.5}$ is missing at monitoring locations. We also performed a \emph{spatial} CV experiment where all observations at each monitor were left out one-monitor-at-a-time. This allowed us to evaluate the model's ability to perform \emph{spatial} interpolation to estimate PM$_{2.5}$ at locations without monitors. We quantified the performance of different methods using the following statistics: prediction root-mean-square error (RMSE), 95\% coverage probability of the posterior intervals (PI), average posterior standard deviation (SD), and R$^2$. R$^2$ and RMSE were calculated based on posterior predictive means of the left-out observed  PM$_{2.5}$ concentrations. Posterior prediction intervals were based on the 2.5$^{th}$ and the 97.5$^{th}$ quantiles of the posterior distribution of the two-component mixture distribution in Equation (\ref{eq:Ensemble}).

Recall that in an effort to only use the data once when estimating ensemble weights, inputs to the ensemble model are based on out-of-sample predictions from the CMAQ and the AOD  downscalers. Hence, in the CV experiments, we compared model performance using ensemble inputs either derived from 10-fold CV or spatial CV. We also evaluated the more computationally efficient \emph{two-stage} estimation approach for ensemble weights compared to a \emph{joint} estimation where the ensemble weights are estimated jointly with the Gaussian process parameters. 

Table~\ref{tab:a} gives model performance results for the 10-fold CV experiment, comparing two downscalers using either AOD or CMAQ to the proposed ensemble approach under different ensemble input choices and estimation approaches.  Overall, the ensemble approach resulted in improved out-of-sample predictions. Specifically, using inputs derived from the 10-fold CV, the ensemble model achieved the lowest RMSE and highest R$^2$. The decrease in posterior prediction SD is particularly significant (about 30\% reduction), while maintaining the proper coverage.  Using spatial CV predictions as inputs resulted in slightly worse performance, likely due to the spatial interpolation that may introduce additional uncertainty. Using a two-stage estimation approach resulted in a small negligible reduction in prediction performance compared to joint estimation. Table~\ref{tab:b} provides performance comparisons from the spatial CV experiment. While results show similar trends as in the 10-fold CV experiment, we find the improvement of the ensemble approach over separate models tends to be better, suggesting the ensemble approach is particularly beneficial for spatial interpolation compared to using only CMAQ or only AOD. 

%% Insert Tables here
\renewcommand{\arraystretch}{1.5}
\begin{table}[ht]
\centering
\caption{Prediction performance for daily PM$_{2.5}$ concentrations in 10-fold cross-validation (CV) comparing ensemble averaging with downscalers using either satellite-derived aerosol optical depth (AOD) or numerical model (CMAQ) simulation. Ensemble inputs were derived from either 10-fold or leave-one-monitor-out (spatial) CV.}\label{tab:a}
\begin{tabular}{ |c|c|c|c|c|c|c| } 
\hline
& && Coverage & Average & \\
Method & Estimation  & RMSE & of 95\% PI& Posterior SD &  R$^2$\\
\hline
\hline
AOD Downscaler & -- &  3.40 & 94.07 & 3.30 & 0.78 \\ 
\hline
 CMAQ Downscaler & --  & 3.14 & 95.05 & 3.28 & 0.81 \\ 
\hline
 Ensemble (10-fold CV input)& Joint & 2.99 & 97.14 & 2.40 & 0.83 \\ 
\cline{2-6}
 & two-stage & 3.01 & 97.08 & 2.49 & 0.82 \\
\hline
 Ensemble (spatial CV input)& Joint & 3.13 & 97.37 & 2.58 & 0.81 \\ 
\cline{2-6}
 & two-stage &  3.14 & 97.33 & 2.62 & 0.81 \\ 
\hline
\hline
\end{tabular}
\\
\end{table}

\begin{table}[h]
\centering
\caption{Prediction performance for daily PM$_{2.5}$ concentrations in leave-one-monitor-out (spatial) cross-validation (CV) comparing the ensemble method with downscalers using either satellite-derived aerosol optical depth (AOD) or numerical model (CMAQ) simulation. Ensemble inputs were derived from either 10-fold or leave-one-monitor-out (spatial) CV.}\label{tab:b}
\begin{tabular}{ |c|c|c|c|c|c|c| } 
\hline
& && Coverage & Average & \\
Method & Estimation  & RMSE & of 95\% PI& Posterior SD & R$^2$\\
\hline
\hline
AOD Downscaler & -- & 3.45 & 94.25 & 3.39 & 0.77 \\ 
\hline
CMAQ Downscaler & -- & 3.33 & 95.32 & 3.45 & 0.78 \\ 
\hline
Ensemble (10-fold CV input)& Joint & 2.99 & 96.80 & 2.38 & 0.83 \\ 
\cline{2-6}
& two-stage & 3.02 & 96.67 & 2.41 & 0.82 \\
\hline
Ensemble (spatial CV input)& Joint & 3.15 & 97.15 & 2.55 & 0.81 \\ 
\cline{2-6}
&  two-stage & 3.16 & 97.19 & 2.53 & 0.81 \\ 
\hline
\hline
\end{tabular}
\\\end{table}

\begin{figure}[h]
\centering
\scalebox{.7}{\includegraphics[trim={2cm 13.7cm 5cm 4.2cm}, clip]{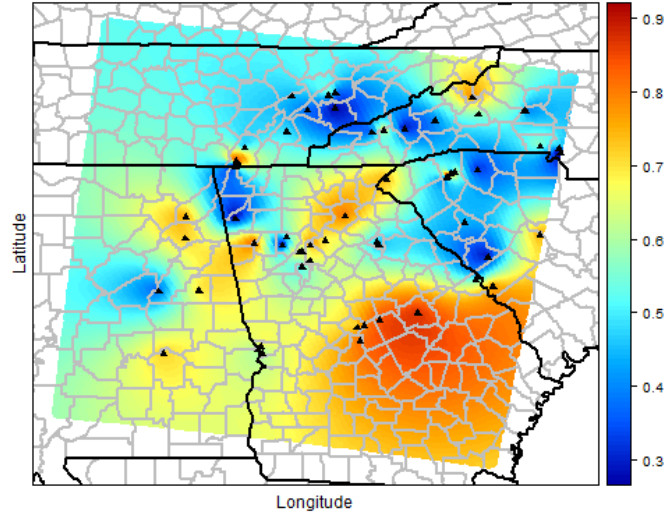}}
\caption{Spatially interpolated ensemble weights for predictions from the Community Multiscale Air Quality (CMAQ) downscaler at 1 km $\times$ 1 km resolution.}
\label{fig:Ensemble}
\end{figure}

To further illustrate the uses of our ensemble approach, we spatially kriged the weight
estimates from the 10-fold CV experiment to areas without monitoring locations at a finer spatial resolution of 1 km $\times$ 1 km across the Southeastern US. We utilized 10,000 MCMC iterations with a burn-in of 5,000 and thinning of 4 to create posterior predictive means of the ensemble weights at each grid cell. Figure ~\ref{fig:Ensemble} clearly demonstrates the need for spatially varying weights due to CMAQ receiving a higher assigned weight value for the mixture model in certain areas, whereas AOD receives higher weights in more rural areas but also close to some urban centers across the study time period.

%%Ensemble figure here

\begin{figure}[h]
\centering
\scalebox{.7}{\includegraphics[trim={3cm 11.4cm 1cm 4.8cm}, clip]{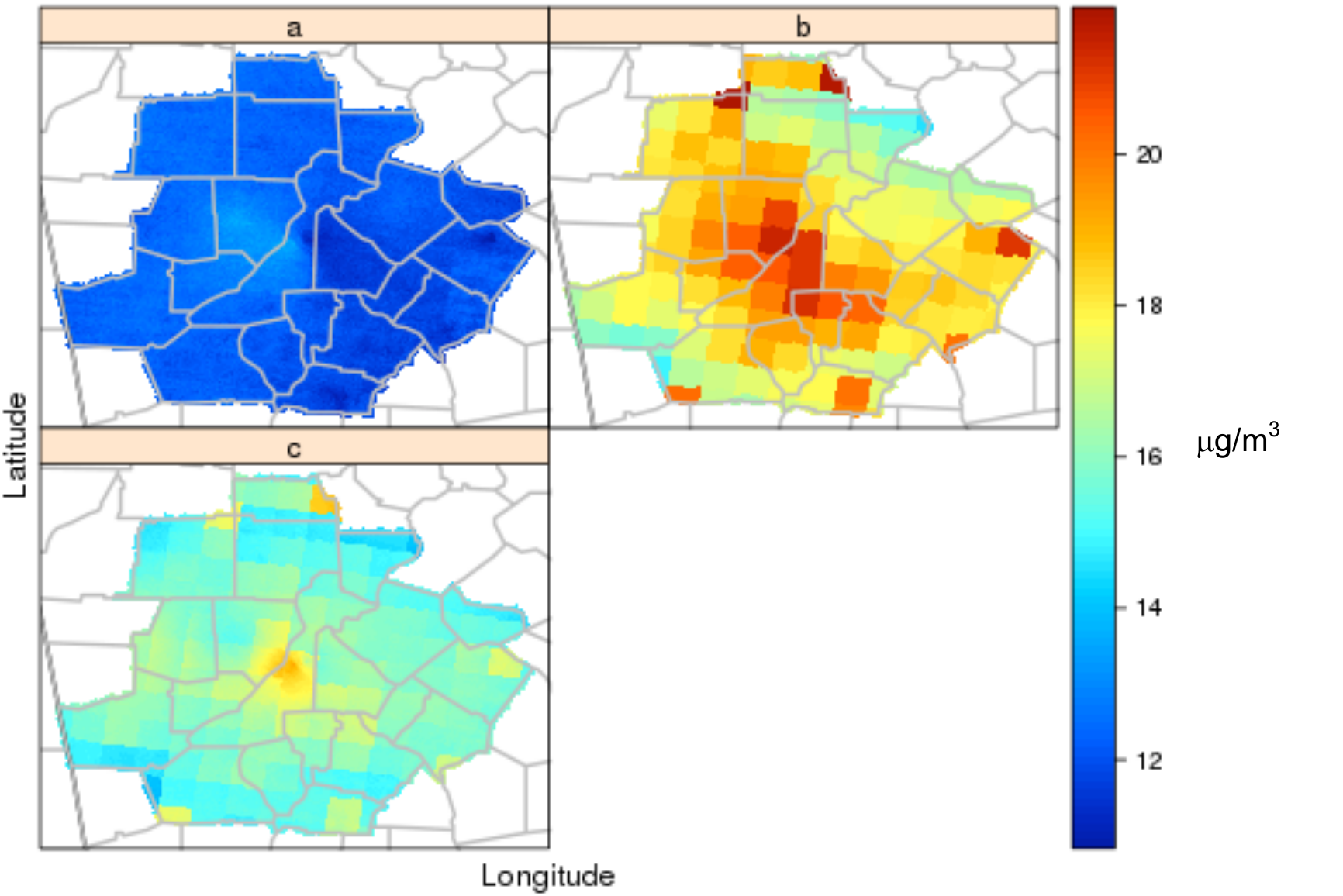}}
\caption{Daily estimates of PM$_{2.5}$ concentrations on March 26, 2015 in the 20-county metropolitan Atlanta, GA area using estimates from a) the AOD downscaler, b) the CMAQ downscaler, and c) the ensemble method.}
\label{fig:sepest}
\end{figure}

While we investigate ways to estimate PM$_{2.5}$ in the Southeastern US, we want to focus on a particular area within that region that has varying weights for the mixture model. PM$_{2.5}$'s environmental health effects are well-documented in Atlanta, GA [\cite{Lungs_Alhanti}; \cite{Lungs_Gass}]. 
To that end, we center our data analysis within the 20-county metropolitan Atlanta, GA area. We aim to contrast results from the two individual data sources with our results from the combined, ensemble method. This Atlanta region contains 16,063 AOD grid cells and 143 CMAQ grid cells, as seen in Figure \ref{fig:sepest}a and Figure \ref{fig:sepest}b, respectively. 

%%sepest figure here

Figure \ref{fig:sepest} demonstrates the applicability of the ensemble approach for a single day. The 
20-county metropolitan Atlanta area has 9 AQS monitors, but the ensemble approach, combined with spatial kriging and interpolation, allows us to extend the use of weights beyond areas with monitors to obtain posterior predictive mean PM$_{2.5}$ concentrations across a wider swath of land. Figure \ref{fig:sepest}b, the CMAQ downscaler results, starkly differs from Figure \ref{fig:sepest}c, the ensemble averaged results, in terms of smoothness. On this particular day, the AOD downscaler predicts lower PM$_{2.5}$ concentrations over Atlanta than the CMAQ downscaler (Figure \ref{fig:sepest}a and Figure \ref{fig:sepest}b). The ensemble approach leads to an average of the AOD and the CMAQ downscaler predictions and, thereby, allows for depictions of seamless PM$_{2.5}$ estimates between neighboring spatial fields for which CMAQ alone does not have the complexity. 

\begin{figure}[h]
\centering
\scalebox{.7}{\includegraphics[trim={3cm 11.5cm 1cm 5cm}, clip]{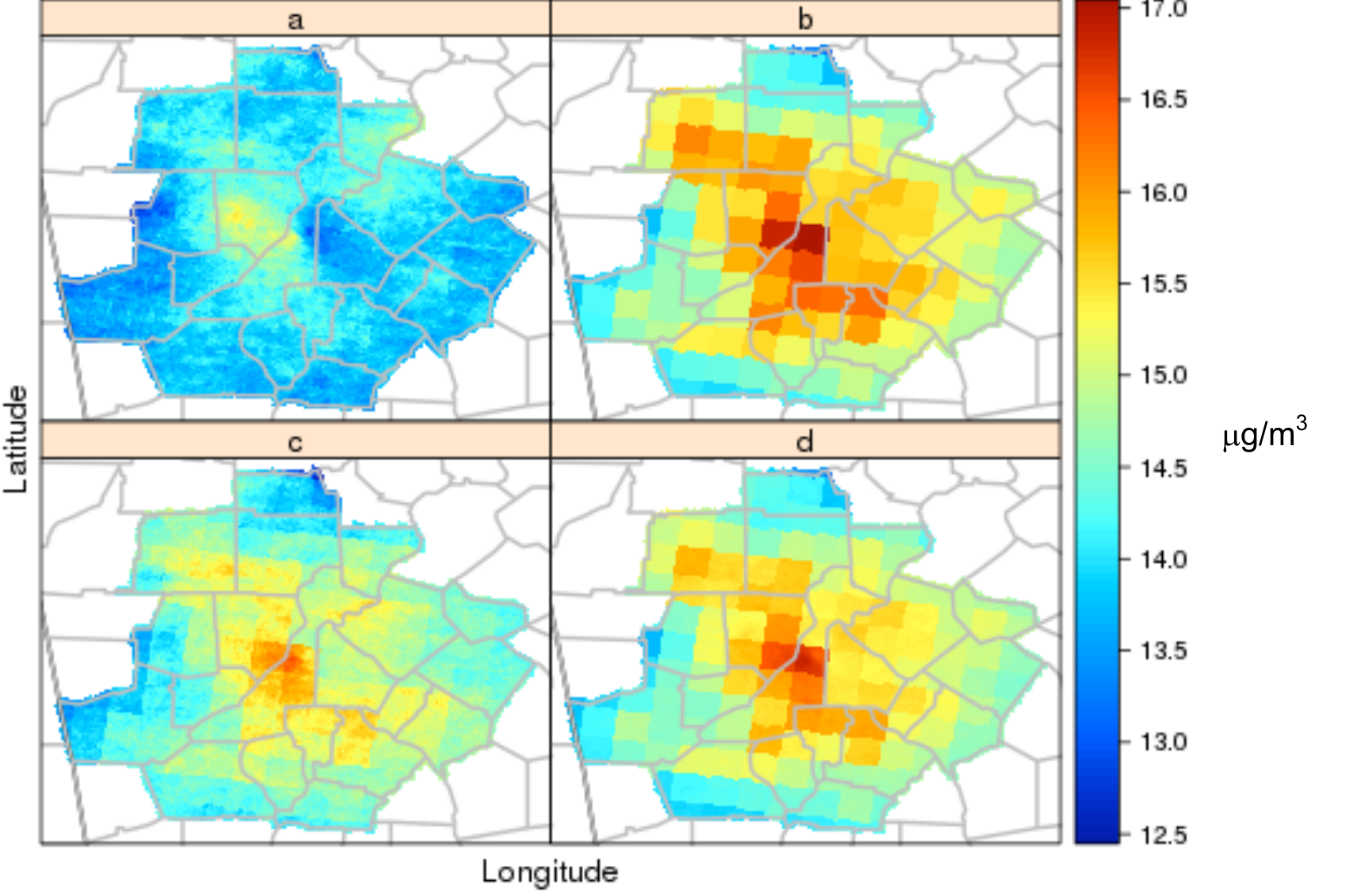}}
\caption{Posterior averages of PM$_{2.5}$ concentrations across 2013-2015 in the 20-county metropolitan Atlanta, GA area based on a) the AOD downscaler, b) the CMAQ downscaler, c) the ensemble method for days in the three-year time period where AOD is observed, and d) the ensemble method for all days in the three-year time period.}
\label{fig:AllPost}
\end{figure}

Figure 5 displays the long-term 3 year PM$_{2.5}$ concentration estimates over Atlanta from the AOD downscaler (Figure \ref{fig:AllPost}a), the CMAQ downscaler (Figure \ref{fig:AllPost}b), and ensemble averages restricted to days when AOD was observed (Figure \ref{fig:AllPost}c) or across all days (Figure \ref{fig:AllPost}d). The combination of information from the AOD downscaler and CMAQ downscaler permits more granularity in the maps on both a daily level (Figure \ref{fig:sepest}c) and when averaging across days where AOD is observed (Figure \ref{fig:AllPost}c). This finer resolution on a daily level or on days with observed AOD will aid in acute environmental health effect analyses. However, in Figure \ref{fig:AllPost}d, the predictions from the CMAQ downscaler dominate, likely due to the large number of missing AOD in this region over time (about 61\%).  

%Figure allpost here

%%%%%%%%%%%%%%%%%%%%%%%%%%%%%%%%%%%%%%%%%%%%%%%%%%%%%%%%%%%%%%%%%%%%%%%%
\section{Discussion}
%%%%%%%%%%%%%%%%%%%%%%%%%%%%%%%%%%%%%%%%%%%%%%%%%%%%%%%%%%%%%%%%%%%%%%%%
%\subsection{Strengths}
Instead of relying solely upon numerical CTM simulations or satellite data to perform data fusion, the proposed Bayesian ensemble model averaging framework allows us to incorporate both sources of information and harness their collective predictive power. Another advantage of the ensemble approach entails accounting for differences in spatial resolution between different gridded data because CTM and satellite data are first calibrated to the point-level using monitoring data via statistical downscaling. Finally, in our PM$_{2.5}$ application, the ensemble approach also naturally accounts for the missing values in satellite retrievals. Specifically, when satellite AOD is missing, ensemble weights for different inputs can be reweighted among available inputs. This differs from existing approaches where AOD needs to be imputed before being used as a predictor for PM$_{2.5}$, increasing computational burden and introducing another source of prediction uncertainty. Although here we focus on ambient air pollution for the application of this method, the approach is also highly relevant to the estimation of other environmental exposures (e.g. temperature, precipitation) that utilize information from both satellite imagery and numerical model simulations. 

Several extensions of the proposed method warrant additional investigations. First, ensemble modeling can be generalized to consider multiple sources of information. For example, one can consider a model only driven by fine-scale land use variables with AOD missing. Specifically, the two-component mixture model utilized here can be extended to have multiple weights (i.e. more than two) that are estimated with a multinomial latent variable with probabilities following a Direchlet prior distribution.  In the air pollution application, this may include (1) CTM simulations driven by different assumptions on emission levels and pollution composition for each emission source, (2) multiple satellite parameters that may inform different characteristics of aerosol, and (3) AOD retrievals from different satellites. We modeled spatially-varying weights largely due to the ability of satellite-retrieved AOD to predict PM$_{2.5}$ over large areas and the error in CMAQ simulation being likely to exhibit spatial heterogeneity. Another extension of the ensemble method is to allow weights to depend on spatial and temporal covariates (e.g. land use and meteorology). This may further improve PM$_{2.5}$ prediction and provide insights into factors associated with when the CMAQ and AOD downscalers are ineffective.

%%%%
%Bibliography
%%Alphabetical for Biometrics set-up

%From Biometrics guide:
%References to published literature should be quoted in the text by author and date, e.g., Draine (1978) or (Begelman,
%Blandford, and Rees, 1984). Where more than one reference is cited having the same author(s) and date, the letters
%a,b,c,. . . should follow the date, e.g., Smith (1988a), Smith (1988b), etc. For papers with exactly three authors, the ?rst
%time the paper is cited, all author names should be used, e.g., Begelman, Blandford, and Rees (1984), and subsequent
%citations should use \et al.," e.g., Begelman et al. (1984). For papers with more than three authors, the ?rst author
%name plus \et al." should always be used. In the bibliography list, all authors should be retained.
\bibliographystyle{imsart-nameyear}
\bibliography{aoasdraftbib}

\end{document}